\documentclass[twocolumn,aps,longbibliography]{revtex4-2}

\usepackage{graphicx}
\usepackage{amsmath}
\usepackage{amssymb}
\usepackage{amsthm}
\usepackage{bm}
\usepackage{xcolor}
\usepackage[colorlinks=true,citecolor=blue,linkcolor=magenta]{hyperref}
\usepackage{braket}
\usepackage{booktabs}
\usepackage{algorithm}
\usepackage{algpseudocode}

\newcommand{\HH}{\mathcal{H}}

\begin{document}

\title{Classical Verification of Quantum Advantage\\
via Clifford Obfuscation}

\author{Bin Yan}
\affiliation{Visa Research, Visa Inc., Foster City, CA 94404}

\date{\today}

\begin{abstract}
Demonstrating quantum advantage in a manner that can be independently verified by classical means remains one of the most pressing open problems in quantum computing. Here we propose a new heuristic approach for constructing quantum circuits that are hard to classically simulate yet whose output distributions can be efficiently verified classically. Our approach is based on \emph{Clifford circuit obfuscation}. This obfuscation scheme hides the Clifford structure and injects a controlled and rapidly growing non-stabilizer resources that resist known classical attack strategies including reverse engineering and direct simulation. Our protocol is heuristic but is supported by both numerical and theoretical evidence. This work provides a new avenue toward classically verifiable quantum advantage that avoids the stringent implementation requirements of known approaches.
\end{abstract}

\maketitle

\section{Introduction}\label{sec:intro}

The question of whether quantum computers can perform tasks intractable for any classical machine has moved from a theoretical curiosity to an experimental frontier~\cite{Arute2019,Zhong2020,King2023SpinGlass,barron2026observable,leviatan2026resolving,Martiel2026Sampling}. Yet a central challenge has emerged alongside this progress: can claimed quantum advantage be independently verified by classical means, without access to a second quantum computer?~\cite{Hangleiter2026} This verification
problem is a practical prerequisite for building trust in quantum
hardware as devices scale, and for establishing quantum computing as a technology whose outputs can be audited independently.

In principle, verification is straightforward for sufficiently powerful quantum computers, e.g., the output of Shor's algorithm can be checked classically in polynomial time, and more generally any \textsc{NP} certificate admits efficient classical verification.
In practice, however, this resolution is not available in the near term. Similarly, protocols based on cryptographic primitives~\cite{Brakerski2021,Mahadev2018,Kahanamoku2022Classically} provide rigorous proofs of quantumness but require circuit depths and coherence properties well beyond current hardware. What is needed is a verification protocol compatible with the shallow, structureless circuits that near-term devices can reliably execute.

Quantum sampling tasks are a natural candidate~\cite{BremnerJozsaShepherd2011,Boixo2018,Hangleiter2023Computational,Aaronson2011}. In random circuit
sampling (RCS), a processor applies a pseudo-random circuit and
outputs bitstrings sampled from the resulting distribution~\cite{Boixo2018,Bouland2019,Hangleiter2023Computational}, a task that is classically hard to simulate under plausible complexity assumptions. Verification has relied on cross-entropy benchmarking (XEB), which estimates the overlap with the ideal output distribution --- but computing that ideal distribution is itself classically intractable in the regime where quantum advantage is claimed, making verification indirect and assumption-dependent~\cite{Hangleiter2026}. A more recent result by Google offers a partial remedy by exploiting constructive interference near the edge of quantum ergodicity to expose verifiable output structure~\cite{Google2025}, but checking this structure at scale still requires a second quantum device, and, like most near-term advantage claims, rests on heuristic hardness assumptions.

The structureless nature of random circuit sampling at the same time makes it classically hard to verify. This dilemma suggests that more robust and practical resolutions might be to take a step back and seek heuristically secure protocols: schemes that engineer a ``trapdoor''---a planted secret---into the circuit to enable efficient classical verifications. Peaked circuit sampling~\cite{AaronsonZhang2024,Gharibyan2025}, for instance, pursues the idea along this line by concentrating the output distribution on a hidden peak bitstring, reducing verification to a simple statistical spot-check. However, whether such circuits can be efficiently constructed remains largely unknown. As quantum computers move beyond the NISQ era, with early error-correction capabilities emerging~\cite{Preskill2025Beyond,Bluvstein2024Logical,GoogleSurface}, efficient schemes for verifying and benchmarking their performance become increasingly important~\cite{GrandClallenge}.

\begin{figure*}[t!]
    \centering
    \includegraphics[width=\linewidth]{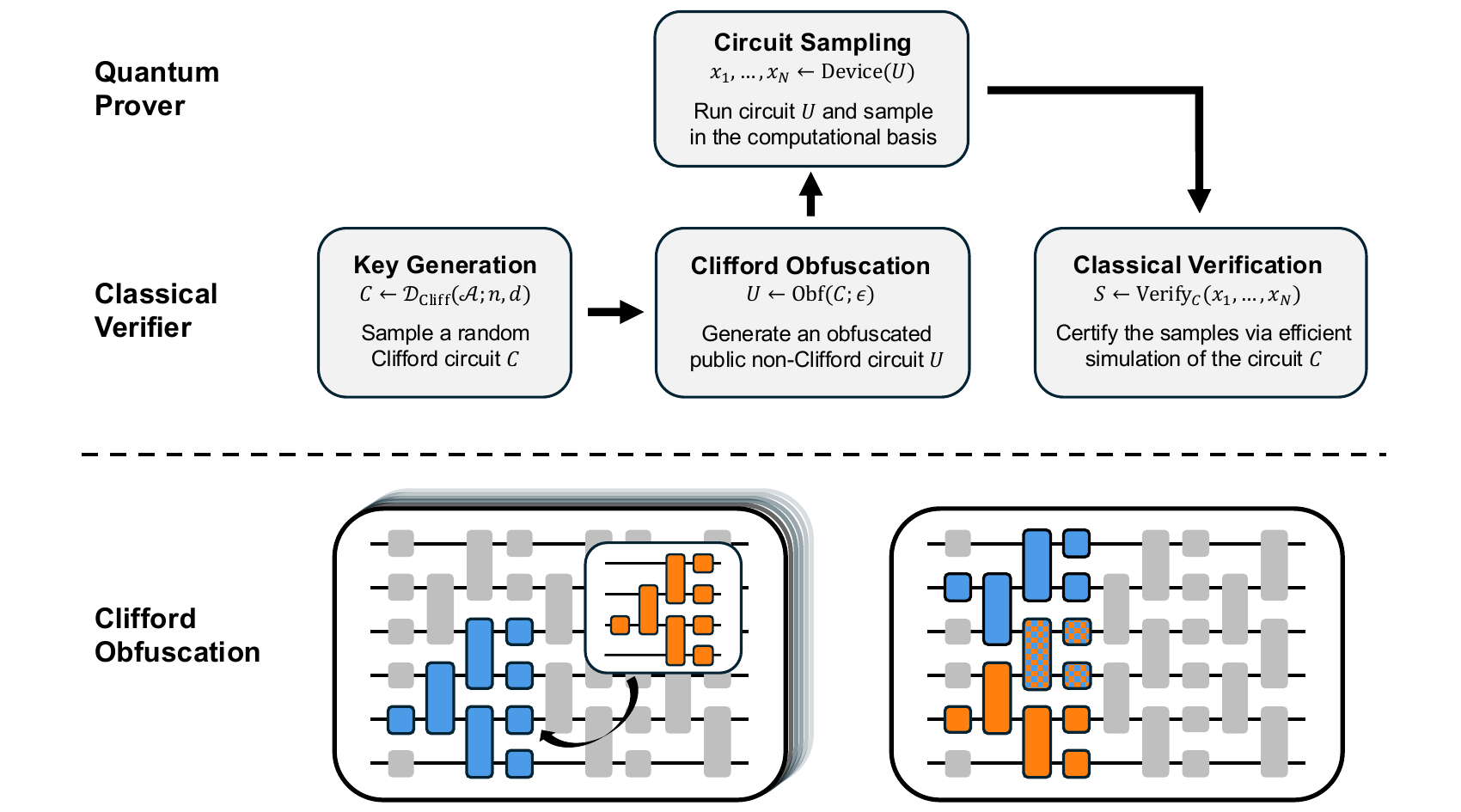}
    \caption{Upper panel: The verification protocol incorporates Clifford obfuscation into random circuit sampling. Lower left: The obfuscation scheme consists of repeated patch updates. At each step, a local patch, a finite convex region (blue) within the circuit layout, is selected and replaced by a patch of the same shape (orange) that realizes the same unitary (or approximately with high fidelity) but has different gate parameters. This process is repeated until the entire circuit is covered. Lower right: Crucially, the updated patch regions overlap, allowing local information in the circuit to be scrambled. For example, after two overlapping patches are updated, neither individual patch realizes the same Clifford unitary as it did in the original circuit.}
    \label{fig:circuit}
\end{figure*}

Our proposal to fill this gap is motivated by the same spirit but takes a distinct approach. Specifically, we take a Clifford circuit as a trapdoor and progressively obfuscate it into a circuit that is designed to appear random under local diagnostics and hard to simulate directly. However, for a verifier who knows the original Clifford circuit, the probabilities of sampled bitstrings can be computed efficiently~\cite{GottesmanKnill,AaronsonGottesman2004}, making it possible to perform direct verification. Clifford circuits are a natural starting point for this task: they admit a well-understood resource theory of non-Clifford content through T gates and magic
monotones~\cite{Veitch2014,HowardCampbell2017,BravyiGosset2016,PashayanWallmanBartlett2015}. The gap between Clifford circuits and universal quantum computation is precisely the non-Clifford resource gap, and our protocol exploits this gap in a controlled way.

We provide strong numerical evidence that the protocol is secure against the two most natural classical attacks: direct simulation of the obfuscated circuit, which is obstructed by the rapidly growing non-Clifford resource; and reverse engineering of the original Clifford circuit, which is obstructed by the effective randomization of the local circuit parameters through scrambling of local information. The protocol requires no cryptographic primitives with deep circuit implementation and no second quantum device, making it well-suited to the near-term verification problem.

\section{Obfuscation Scheme}\label{sec:scheme}
For a given unitary, there are often combinatorially many distinct realizations in terms of parameterized quantum circuits. We leverage this flexibility to compile a Clifford unitary into a parameterized circuit, either exactly or up to a controlled approximation error, such that the resulting parameters do not reveal exploitable structure to a classical adversary attempting to reconstruct the original Clifford description. The compiled circuit is then provided to a quantum prover for random circuit sampling. See Figure~\ref{fig:circuit} for a schematic of the verification protocol. Crucially, the sampling process remains classically verifiable: a verifier who possesses the ``Clifford key'' can efficiently compute the output probabilities.

The basic obfuscation primitive is repeated updating of \emph{local patches} of a parameterized quantum circuit. Represent the circuit in the standard spacetime diagram, where qubits span the spatial dimension and gate executions span the time dimension. The circuit is initialized as containing only random Clifford gates~\footnote{As discussed later, we restrict to Clifford circuits whose output states do not have full support, i.e., are not supported on all computational-basis states.}. As shown in Figure~\ref{fig:circuit}, in each updating step, we identify a  convex local patch that implements a well-defined unitary. We then replace it with a local patch that has exactly the same shape (gate layout) and represents the same unitary (or close to that unitary). However, the gate parameters differ from those in the original patch. The above process is repeated, until the entire circuit diagram is covered by the updated patches at different locations. This general obfuscation scheme is summarized as pseudocode in the following algorithm.

\noindent\begin{minipage}{\linewidth}
\vspace{8pt}
\hrule height 1pt
\vspace{3pt}
\center{Algorithm 1: Clifford Obfuscation \label{alg:obfuscate}}
\vspace{3pt}
\hrule
\vspace{3pt}
\begin{algorithmic}[1]
\Require Circuit ansatze $C(\theta)$, number of update passes $p$
\State Initialize $\theta \leftarrow$ Clifford circuit key
\For{pass $= 1$ \textbf{to} $p$}
    \State Identify a local patch $S$
    \State Update $\theta' \leftarrow \theta$ within $S$ and keep $U_S(\theta) \approx U_S(\theta')$
\EndFor
\State \Return $\tilde{C}(\theta)$
\end{algorithmic}
\hrule
\vspace{8pt}
\end{minipage}

Several remarks are in order.

(i) A crucial part of the obfuscation is that the updated patches are overlapping, and a local point in the circuit diagram might be covered by multiple patches. This ensures that, cf.,~Figure~\ref{fig:circuit}, right, after patch updating, information about the local gate configurations is scrambled across the entire circuit, and the unitaries corresponding to local circuit regions are pushed away from the original Clifford unitaries in those regions. 

(ii) If every patch were updated exactly, the global circuit would remain Clifford-equivalent and no non-Clifford resource would be generated for unitary of the global circuit. Nevertheless, the gate parameters get randomized, and local regimes of the circuit become non-Clifford. We conjecture that it is computationally hard to classically sample such circuits. Note that the conjectured hardness is not from the circuit unitary \emph{per se}, which is Clifford and \emph{can} be efficiently simulated when the circuit is presented as a sequence of Clifford gates. The hardness stems from the assumption that it is classically hard to distinguish the obfuscated Clifford circuit from a generic random circuit, and sampling of the latter is well believed to be classically hard.

\begin{figure*}[t!]
    \centering
    \includegraphics[width=\linewidth]{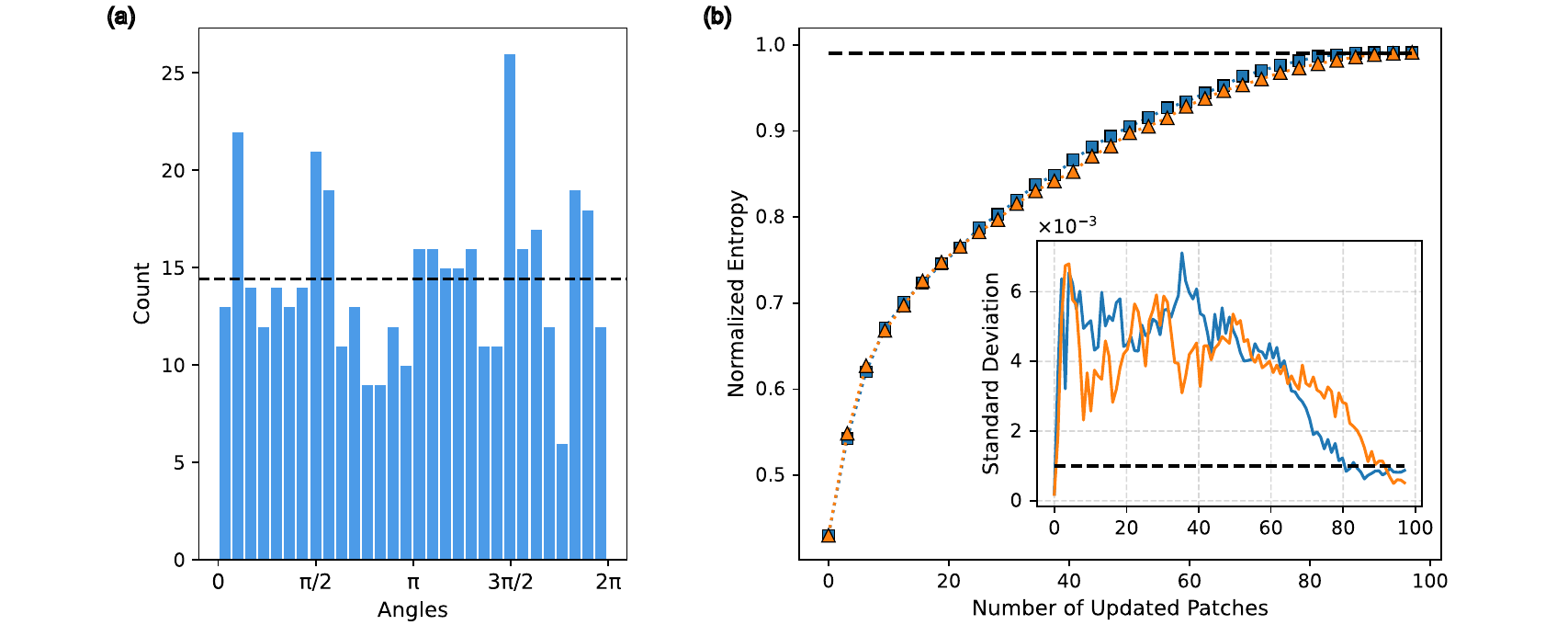}
    \caption{(a) Histogram of the single-qubit rotation angles for a typical $12$-qubit instance after one obfuscation cycle with the simulation details discussed in Algorithm~\ref{alg:obfuscate} and Section~\ref{sec:simulation}. Dashed horizontal line corresponds to the average count. Clifford values of these angles are multiples of $\pi/2$. (b) The normalized entropy of the single qubit rotation angles after obfuscation of the same $12$ qubit system. Data points are obtained by averaging over $10$ simulation runs. The blue square and orange triangles correspond, respectively, to perfect obfuscation (fidelity of the obfuscated circuit is larger than $0.9999$ w.r.t. to original Clifford circuit) and finite-fidelity obfuscation, with average fidelity~$\sim 0.6$. Inset shows the standard deviations of the corresponding data in the main figure. The horizontal dashed lines indicate the expected values if the angles were independently sampled from the uniform distribution.}
    \label{fig:entropy}
\end{figure*}

(iii) It is also important to stress that the ``indistinguishability'' reasoning is false for quantum adversaries: Given quantum oracle access to the circuit, it is efficient to reconstruct the Clifford circuit~\cite{Low2009Learning,Chen2025Stabilizer}. Of course, this does not necessarily falsify verification of quantum advantage, as the reconstruction involves quantum computers. Nevertheless, to make the protocol stronger and quantum-safe, we can selectively perform imperfect patch updating with controlled residual mismatches. This would result in a genuine non-Clifford circuit with finite fidelity with respect to the original Clifford circuit. Non-Clifford circuits are believed to be generally hard to simulate, e.g., requiring a computational resource that scales exponentially as the number of $T$ gates within the circuit. As demonstrated below, the amount of non-Clifford component injected into the circuit through imperfect patch updating saturates \emph{exponentially} fast with the circuit infidelity. Crucially, the finite fidelity of the obfuscated circuit does not invalidate the Clifford key. For example, in random circuit sampling, comparing samples generated by the obfuscated circuit with the ideal sampling probabilities of the original Clifford circuit would result in a reduction in XEB that is proportional to the circuit fidelity.

(iv) Since each local patch is finite-sized, updating a single patch incurs only constant computational cost. Moreover, a number of patches that scales linearly with the circuit size is sufficient to cover the entire circuit layout. Consequently, the overall cost of the obfuscation protocol scales linearly with the size of the circuit. In this work, simulations were restricted to relatively small system sizes for validation, as evaluating the full circuit fidelity and determining lower bounds on T-gate counts require explicit manipulation of full unitary matrices. These evaluations are not part of the obfuscation protocol itself.

In the following, we discuss a concrete realization of the obfuscation and present numerical evidence supporting the hardness assumptions. 

\section{A Case Study}\label{sec:simulation}

The high-level obfuscation procedure in Algorithm~\ref{alg:obfuscate} does not specify the methods for selecting local patches and updating the gate parameters. A good practice for designing such procedures is to randomize the local gate parameters as much as possible to hide the Clifford structure, and possibly inject non-Clifford resources. As a concrete example, we used the following circuit ansatz and patch updating procedures for numerical simulations:

\begin{enumerate}
\item \emph{Circuit parameterization}: We consider a brickwork architecture on $n$ qubits with depth $d$. Each layer consists of a single-qubit rotation sublayer followed by two fixed entangling CZ sublayers,
\begin{equation}
    U_l = \mathrm{CZ}_{\mathrm{odd}}\,\mathrm{CZ}_{\mathrm{even}} \bigotimes_{q=1}^{n} R_q^{(l)},
\end{equation}
where $R_q^{(l)} = R_x(\gamma_q^{(l)})R_z(\beta_q^{(l)})R_x(\alpha_q^{(l)})$
is a general single-qubit Euler rotation. The full circuit unitary is $U = U_d \cdots U_1$. The original Clifford circuit is obtained by restricting the Euler angles to Clifford-compatible values (multiples of $\pi/2$), making every single-qubit gate Clifford and the whole circuit efficiently classically simulable~\cite{GottesmanKnill,AaronsonGottesman2004}.

\item \emph{Patch selection}: For each spacetime coordinate $(q, l)$ associated with a single qubit gate, we select a local patch as the forward light cone of depth $L$, that is, the set of gates whose output can be influenced by a perturbation at $(q,l)$ within $L$ layers (see Figure~\ref{fig:circuit}, lower left, for an illustration). As an optimization, we remove the final CZ gate sublayer at the output of each full light cone, since it carries no trainable parameters and only enlarges the local Hilbert space. Patch updates were then performed for all such light cones over the relevant values of $q$ and $l$, until all such light cone patches in the circuit layout are updated. This process can be repeated for multiple cycles. In this work, the simulations were performed for one cycle.

\item \emph{Patch updating}: For each patch $S$, let unitary $V_S$ the
corresponding local restriction of the circuit \emph{before} updating, and let $U_S(\theta)$ be a unitary that corresponds to a circuit patch with the same gate layout as the local patch $S$, and with free single-qubit parameters $\theta$. We then train the parameterized patch $U_S(\theta)$ to make it close to the target unitary $V_S$. This is achieved through minimizing the normalized Hilbert--Schmidt
loss
\begin{equation}
    \mathcal{L}_S(\theta) = 1 - \frac{|\mathrm{tr}(V_S^{\dagger}U_S(\theta))|^2}{\dim(\HH_S)^2}.
\end{equation}
An update is accepted only if $\mathcal{L}_S < \epsilon_{\mathrm{tol}}$. The acceptance threshold $\epsilon_{\mathrm{tol}}$ is the primary control parameter governing the fidelity of the obfuscated circuit and scales with the infidelity, $\epsilon$, of the final circuit w.r.t. the original Clifford circuit as $\epsilon_{\mathrm{tol}} \sim \epsilon / M$, where $M$ is the total number of updated patches.
\end{enumerate}

\section{Computational Hardness}\label{sec:hardness}
The proposed protocol is empirical and does not currently admit a formal complexity-theoretic security proof. The security instead rests on two distinct hardness arguments, corresponding to the two most natural attack strategies available to a classical adversary.

\subsection{Attack I: Reverse Engineering}

The first attack attempts to recover the original Clifford circuit from
the obfuscated one, thereby reconstructing the verifier's classical
reference. A successful adversary would need to identify the underlying Clifford structure despite the accumulated non-Clifford perturbations. We argue that this is computationally infeasible on the grounds that the obfuscation protocol effectively randomizes the single-qubit gate parameters beyond any distinguishable structure.

To quantify this, we compute the (normalized) entropy of the distribution of the single-qubit Euler angles in the obfuscated circuit --- the continuous angles parameterizing each single-qubit gate --- as a function of the number of updated patches. We partition the range of angles, $[0, 2\pi]$, into a finite number of bins, and compute the probability $p_i$ for the angles to be in the $i$-$th$ bin. The normalized entropy is defined as

\begin{equation}
    E = - \left(\sum_i p_i\ln{p_i}\right)/H_{\rm unif},
\end{equation}
where the normalization constant $H_{\rm unif}$ is the Shannon entropy of the uniformly distribution evaluated with the same number of bins. 

\begin{figure*}[t!]
    \centering
    \includegraphics[width=\linewidth]{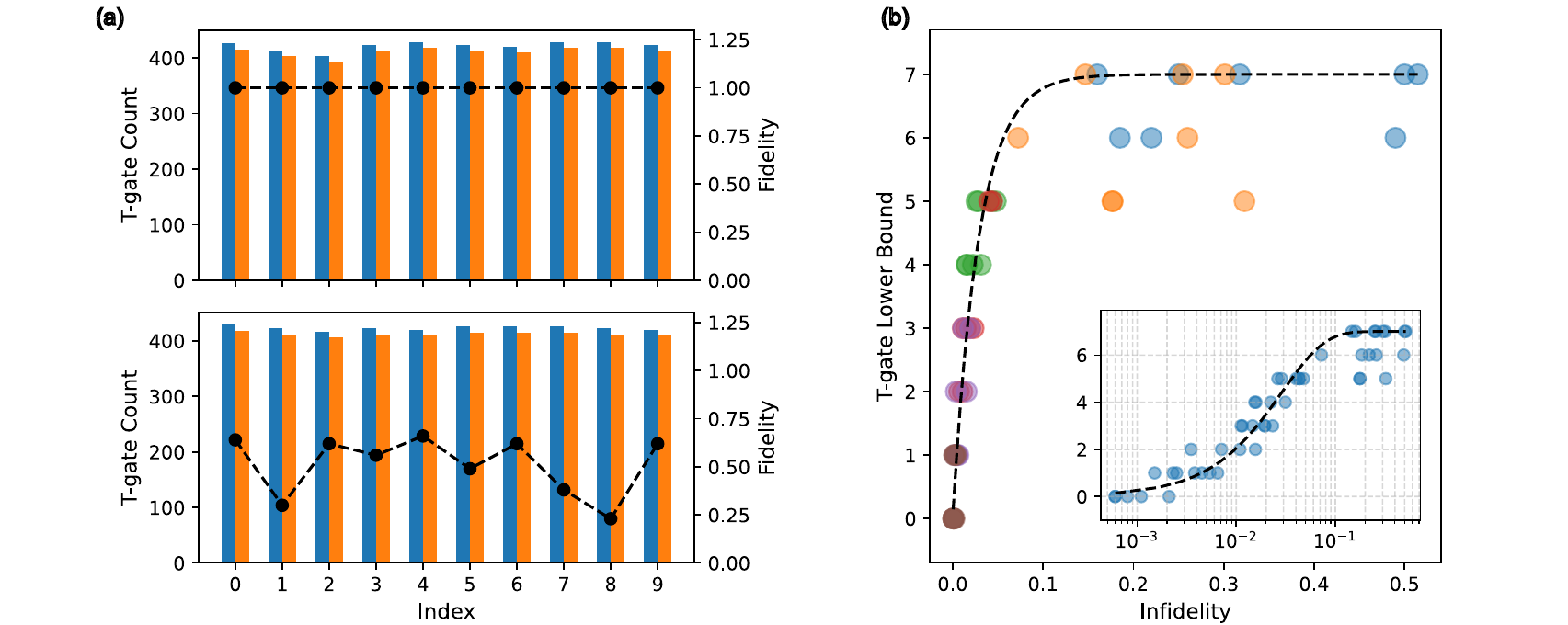}
    \caption{(a) T gate counts before (blue bars) and after (orange bars) the PyZX reduction, where the index labels various runs of the simulation. The top and bottom panels, respectively, correspond to data from circuits after perfect obfuscation (fidelity of the obfuscated circuit is larger than $0.9999$ w.r.t. the original Clifford circuit) and obfuscation with finite fidelity (averaged fidelity $\sim 0.6$). The fidelity is marked by the black dots (plotted against the right vertical axis). (b) The T gate lower bound evaluated from the unitary stabilizer nullity at various infidelities of the trained circuit w.r.t. to the original Clifford unitary. The dashed line represents the best fit to a saturated exponential function~\eqref{eq:exponential} with a single fitting parameter (exponential exponent). Data points with the same color are generated using the same set of hyperparameters controlling the final circuit fidelity. Inset shows the same plot with the x-axis in logarithmic scale.}
    \label{fig:fidelity}
\end{figure*}

As shown in Figure~\ref{fig:entropy}~(b), the entropy saturates rapidly to that of the uniform distribution, indicating that the parameter distribution becomes indistinguishable from uniform at the \emph{lowest order}, though hidden higher order correlations might be still extractable. Crucially, this saturation is observed for both perfect and imperfect patch updating, with no significant visible difference between the two curves. This demonstrates that the randomization is a structural consequence of the obfuscation protocol itself, not an artifact of the injected non-Clifford resources from imperfect patch updating. Figure~\ref{fig:entropy}~(a) shows the parameter histogram for a representative instance for perfect patch updating. Together, these results imply that an adversary inspecting the obfuscated circuit gains essentially no \emph{local} information about the original gate angles, providing a significant barrier for reverse engineering the Clifford circuit.

This randomization of parameters is a direct consequence of the light-cone updates: many inequivalent parameter settings achieve the same local unitary action, and the optimizer selects one that is generically far from the original Clifford values. 

On a global circuit level, as an independent check, we applied PyZX~\cite{KissingerVanDeWetering2020,KissingerVanDeWetering2019PyZX}, a circuit optimization package that reduces T-gate count via ZX-calculus rewriting rules~\cite{CoeckeKissinger2017}, to the obfuscated circuits. If residual Clifford structure were efficiently recoverable, ZX-based simplification would substantially reduce the T-gate count. Figure~\ref{fig:fidelity}~(a) shows that the reduction is negligible, providing further evidence against efficient recovery of the Clifford skeleton. Again, the failure of ZX-calculus there appears to have no visible correlation with the fidelity of the obsucated circuit.

\subsection{Attack II: Direct Simulation}

The second attack attempts to simulate the obfuscated circuit directly, without access to the original Clifford description. This attack is particularly relevant because local obfuscation alone does not necessarily guarantee direct simulation hardness. For example, Ref.~\cite{Kremer2026Efficient} shows that identity circuits of the form $U^\dagger U$ obfuscated using a similar local obfuscation scheme, as considered in Ref.~\cite{Gharibyan2025}, can remain vulnerable to matrix-product-operator simulations. 

The present protocol is designed to avoid this weakness by distributing non-Clifford structure throughout the circuit. Even in the exact obfuscation scenario, where the obfuscated circuit is globally a Clifford circuit, direct layer-by-layer simulation of the circuit would face significant hurdles. To see the intuition, consider cutting the circuit at an intermediate time slice. The subcircuit before the cut generally implements a non-Clifford unitary, with an effective T gate count proportional to the total size of the local patches intersected by the cut. For the patch geometries considered here, this total patch size scales linearly with the number of qubits. This suggests that tensor-network simulations face an extensive non-Clifford obstruction, although we do not claim this as a formal lower bound on tensor-network simulation complexity.

To further strength the protocol against the attack of direct simulation, we can deliberately reduce the fidelity between the obfuscated unitary $U_{\mathrm{obf}}$ and the original Clifford circuit $U_{\mathrm{C}}$, defined as
\begin{equation}\label{eq:fidelity}
    F = |\mathrm{tr}(U_{\mathrm{C}}^{\dagger}U_{\mathrm{obf}})|^2/d^2,
\end{equation}
where $d$ is the dimension of the unitary. This reduction enables the controlled injection of non-Clifford content into the circuit, which is a key resource governing classical simulability. However, fidelity alone is not a reliable measure of non-Cliffordness. For example, if $U_{\mathrm{obf}} \equiv U_{\mathrm{C}}\cdot X$, where $X$ is a single qubit NOT gate, then the fidelity with $U_{\mathrm{C}}$ is zero, even though $U_{\mathrm{obf}}$ remains a Clifford unitary. Nevertheless, fidelity serves as an efficiently tunable parameter within the obfuscation protocol. In the randomized patch updating scheme studied here, it exhibits a positive correlation with the underlying non-Clifford resource in the circuit, which we quantify rigorously.

The non-Clifford resource can be concisely captured by the T gate count: any stabilizer-based classical simulator incurs a cost that grows exponentially in the number of T gates~\cite{BravyiGosset2016}. A rigorous lower bound on the T gate count, $T_{\rm lower}$, is provided by the \emph{unitary stabilizer nullity}~\cite{Jiang2023lower}, defined as the codimension of the largest stabilizer group compatible with the action of the unitary $U$ (see the appendix for a brief introduction of this quantity). 
It is important to distinguish the nullity-derived lower bound from the true optimal T gate count. If $T_{\mathrm{optimal}}$ is the minimum T gate count over all equivalent circuit representations and $T_{\mathrm{circuit}}$ is the T gate count in a specific circuit representation, then
\begin{equation}
    T_{\mathrm{lower}} \;\leq\; T_{\mathrm{optimal}} \;\leq\;
    T_{\mathrm{circuit}},
\end{equation}
The true optimal T gate count may be significantly larger than the lower bound.

We computed $T_{\rm lower}$ for the obfuscated circuits as a function of the circuit infidelity $\epsilon = 1 - F$. The results, shown in Figure~\ref{fig:fidelity}~(b), shows that the T gate lower bound grows rapidly and saturates exponentially fast with infidelity, modeled by a saturated exponential function
\begin{equation}\label{eq:exponential}
    T_{\rm lower} = n(1-e^{-\lambda \epsilon}),
\end{equation}
where $n$ is the total number of qubits and $\lambda$ is the only fitting parameter; it depends on the circuit structure and the particular obfuscation  procedure. This scaling implies that even modest infidelity rapidly drives the lower bound toward an extensive value. 

Taken together, the two sets of numerical results draw a consistent picture: the obfuscated circuit is both structurally opaque (resisting reverse engineering) and classically hard to simulate directly. While we do not claim a formal complexity-theoretic proof of security --- such proofs remain out of reach for circuit obfuscation in general --- the evidence from both attack channels strongly supports the practical hardness of the protocol.

\section{Classical Verification}\label{sec:verify}

\subsection{Stabilizer-Support Benchmarking}

Once samples are obtained from the quantum processor, the verifier can test them against the support of the ideal Clifford output state. This provides a simpler benchmark than XEB in the present setting, because the output distribution of a Clifford circuit is flat on an efficiently computable affine subspace and exactly zero outside it. Let $U_C$ denote the hidden Clifford circuit unitary known to the verifier, and let 
$|\psi_C\rangle = U_C |0\rangle^{\otimes n}$
be its ideal output state. Since $|\psi_C\rangle$ is a stabilizer state, its computational-basis support is an affine subspace $S_C \subseteq \{0,1\}^n$. Let $|S_C| = 2^r$ denote the size of the support. The state can therefore be written as 
\begin{equation}
|\psi_C\rangle = \frac{1}{\sqrt{2^r}} \sum_{z\in S_C} \omega(z) |z\rangle , \qquad \omega(z)\in \{\pm 1,\pm i\}, 
\end{equation} 
up to an overall phase convention. Consequently, the ideal measurement distribution is 
\begin{equation}
p_C(z) = 
\begin{cases} 
2^{-r}, & z\in S_C,\\ 
0, & z\notin S_C . 
\end{cases}  
\end{equation} 

The support $S_C$ can be computed efficiently from the stabilizer tableau of the Clifford circuit~\cite{GottesmanKnill,AaronsonGottesman2004}. For a measured bitstring $z$, define the support indicator 
\begin{equation}
\chi_C(z) = 
\begin{cases} 
1, & z\in S_C,\\ 
0, & z\notin S_C . 
\end{cases}  
\end{equation} 
Given $N$ samples $\{z_i\}_{i=1}^N$, we define the stabilizer-support benchmarking~(SSB) statistic
\begin{equation}\label{eq:support_statistic}
s = \frac{1}{N}\sum_{i=1}^{N} \chi_C(z_i).  
\end{equation} 
Thus $s$ is the empirical fraction of measured samples that lie in the support of the verifier's hidden Clifford state. For ideal sampling from the exactly obfuscated Clifford circuit, every output bitstring lies in $S_C$, and hence 
\begin{equation}
s_{\rm ideal}=1.  
\end{equation} 
This is the basic completeness property of the test.

An adversary who does not know the hidden Clifford structure, and therefore has no information about the affine support $S_C$, can do no better than guess bitstrings in $\{0,1\}^n$. For uniformly random samples, each bitstring lands in $S_C$ with probability 
\begin{equation}
q_r = \frac{|S_C|}{2^n} = 2^{r-n}.  
\end{equation} 
Therefore,  the expectation value and the variance of the SSB are
\begin{equation}\label{eq:random_SSB}
\mathbb{E}_{\rm rand}[s] = 2^{r-n}, \qquad \mathrm{Var}_{\rm rand}(s) = \frac{2^{r-n}\left(1-2^{r-n}\right)}{N}.  
\end{equation} 
Equivalently, $Ns$ follows a binomial distribution. Thus, when $r\ll n$, a random sampler produces an exponentially small value of $s$, whereas an honest ideal sampler produces $s=1$. The verifier can efficiently control this separation by choosing Clifford circuits whose output stabilizer states have any desired support dimension $r$. For a uniformly random Clifford circuit, the support dimension $r$ is typically close to $n$, so the random-sampling baseline $2^{r-n}$ need not be exponentially small. Nevertheless, even in the case $r=n-1$, the random baseline is only $1/2$, while the ideal value remains $1$; by taking a sufficiently large number of samples $N$, the variance in equation~\eqref{eq:random_SSB} can be suppressed as $1/N$, allowing the two cases to be distinguished with high confidence.

The same statistic also has a simple interpretation under depolarizing noise. This allows us to quantify the robustness of SSB to typical noise, and provides a simple phenomenological model the SSB statistics for the finite-fidelity obfuscation scenario. Suppose that the implemented output distribution is modeled as a mixture of the ideal Clifford distribution and the uniform distribution, 
\begin{equation}
p_{\eta}(z) = (1-\eta)p_C(z) + \eta\,2^{-n}, \qquad 0\leq \eta \leq 1,  
\end{equation} 
where $\eta$ is the effective depolarizing strength. Then the expected SSB score is 
\begin{equation}\label{eq:depolarizing_SSB}
\mathbb{E}_{\eta}[s] = (1-\eta) + \eta\,2^{r-n}.  \end{equation} 
Thus depolarization reduces the ideal value $s=1$ linearly toward the uninformed-random baseline $2^{r-n}$. More generally, if the experimental distribution has total probability mass $P(S_C)$ on the stabilizer support, then $\mathbb{E}[s]=P(S_C)$ and 
\begin{equation} 
\mathrm{Var}(s) = \frac{P(S_C)\left[1-P(S_C)\right]}{N}. 
\end{equation} 
The derivation of equations~\eqref{eq:random_SSB} and \eqref{eq:depolarizing_SSB}, as well as the efficient computation of $S_C$ from the Clifford tableau, is given in the appendix.

\subsection{Peaked Circuit}

Clifford obfuscation serves as a primitive for hiding Clifford structure and can be combined with other verification approaches. Here, we discuss a variant of our protocol that allows us to verify the sampling directly with a different strategy from stabilizer-support benchmarking. This approach integrates the peaked random circuit idea proposed in~\cite{AaronsonZhang2024}. 

Peaked circuits refer to quantum circuits whose output distributions contain bitstrings with sufficiently large probabilities, i.e., $\sim 1/{\rm poly}(n)$. Hence, the verifier who knows the peaked bitstrings can verify the occurrence of the peaked bitstrings with a polynomial number of samples. However, it remains largely unknown whether such circuits can be efficiently constructed and simultaneously remain hard to classically sample~\cite{Kremer2026Efficient,Gharibyan2025}.

Now consider a \emph{shallow} parameterized circuit $C_{\rm shallow}$ that is classically trained to produce a peaked output distribution. We can construct a composite circuit 
\begin{equation}
    C \equiv C_P \circ C_{\rm Cliff} \circ C_{\rm shallow},
\end{equation}
where $C_{\rm Cliff}$ is a random Clifford circuit, and $C_P$ is a layer of single qubit gates. 

Note that, by construction of the peaked distribution, the output state of $C_{\rm shallow}$ produces a measurable finite expectation value for the product of Pauli-Z operators, $\sigma^{\otimes n}_Z$. The Clifford part maps $\sigma^{\otimes n}_Z$ to another Pauli string $P$. The single qubit gates in $C_P$ are chosen to map all the non-identity single-qubit Pauli operators in $P$ back to $\sigma_Z$. Hence, $C$ preserves the large peak signal of the original peaked circuit $C_{\rm shallow}$ for computational basis measurements. 

Finally, after applying the obfuscation protocol to the entire composite circuit $C$, the obfuscated circuit is sent to the quantum prover to sample. For a sufficiently large (polynomially) number of samples, the verifier can observe the peak bitstring with a high probability. In this setting, the non-Clifford resource of the obfuscated circuit originates from the shallow circuit itself, rather than from imperfections in the patch-updating procedure.

\bigskip
\section{Discussion}\label{sec:discussion}

We have introduced Clifford circuit obfuscation, a heuristic protocol that produces quantum circuits hard to simulate classically while retaining a verifiable relationship to a classically tractable reference. The core mechanism, staggered local patch updating, provides a practical and controllable route to injecting non-Clifford resource, and our numerical experiments demonstrate that this injection is rapid: even modest training infidelity produces circuits with substantial T-gate resource.

We also discussed a variant of the protocol that combines Clifford obfuscation with a shallow peaked circuit. This approach injects non-Clifford resources without sacrificing fidelity of the obfuscated circuit. Similarly, one can compose a new circuit as $C \equiv C_{2} \circ C_{\rm Cliff} \circ C_{1}$, where $C_1$ and $C_2$ are circuits that either permute computational basis states or consist solely of phase gates, such as T gates and two-qubit controlled Z-rotations, which act by adding phases in the computational basis. These operations do not affect the classical simulability of the circuit $C$ when the Clifford key is known, but they introduce non-Clifford resources $C_1$ and $C_2$ that would spread throughout the circuit after obfuscation.

At its present stage, the hardness of the protocol relies on heuristic assumptions rather than formally established complexity-theoretic guarantees. It is therefore an important direction for future work to investigate whether the security of the protocol can be reduced to a known hardness problem, or at least framed within an established complexity-theoretic paradigm. The gate parameters in our parameterization scheme are defined modulo $2\pi$, so they live on a compact, periodic space. Imperfect patch updating with added noise over this modular space is suggestive of ``learning with noise'' scenarios, such as Learning Parity with Noise~(LPN), as the stabilizer-support benchmarking naturally induces a ``parity check''. At the moment, however, this connection is only intuitive. This may be a promising avenue for future research.

Despite these open questions, the present results establish Clifford circuit obfuscation as a promising route to verifiable quantum advantage. At the very least, in the ``trusted experimenter'' scenario, where the quantum prover honestly carries out the sampling test, the protocol offers a practical approach to certify quantum computational advantage beyond simple benchmarking; it is straightforward to implement, affords fine-grained control over the trade-off between hardness and verifiability, and integrates naturally with existing near-term quantum processors.

\bibliography{references}

\clearpage
\onecolumngrid

\appendix

\section{Stabilizer-Support Benchmarking}\label{app:ssb} 

In this appendix we collect the elementary facts about stabilizer states used in the stabilizer-support benchmarking (SSB) test, and derive the expected value and variance of the SSB statistic under uninformed random sampling and under a simple depolarizing-noise model. 

\subsection{Stabilizer-state support in the computational basis} 

Let $\mathcal{P}_n$ denote the $n$-qubit Pauli group. A pure $n$-qubit stabilizer state $|\psi\rangle$ is the unique simultaneous $+1$ eigenstate of a subgroup $\mathcal{S}\subset \mathcal{P}_n$ generated by $n$ independent commuting Pauli operators. Equivalently, any Clifford circuit $C$ acting on $|0\rangle^{\otimes n}$ produces a stabilizer state 
\begin{equation} 
|\psi_C\rangle = C |0\rangle^{\otimes n}. 
\end{equation} 
The stabilizer tableau of $|\psi_C\rangle$ can be updated efficiently under Clifford gates according to the Gottesman--Knill theorem. When measured in the computational basis, a stabilizer state has a particularly simple structure: its support is an affine subspace of $\{0,1\}^n$, and the probability distribution is uniform on that support. Thus there exists an affine subspace 
\begin{equation} 
S_C \subseteq \{0,1\}^n 
\end{equation} 
such that 
\begin{equation}\label{eq:app_stab_support_state}  
|\psi_C\rangle = \frac{1}{\sqrt{|S_C|}} \sum_{z\in S_C} \omega(z)|z\rangle, \qquad \omega(z)\in \{\pm 1,\pm i\},
\end{equation} 
up to a global phase convention. 

We write \begin{equation} 
|S_C| = 2^r, 
\end{equation} 
where $r$ is the dimension of the affine support. The ideal output probability is therefore 
\begin{equation}\label{eq:app_stab_distribution} 
p_C(z) = |\langle z|\psi_C\rangle|^2 = 
\begin{cases} 
2^{-r}, & z\in S_C,\\ 
0, & z\notin S_C . 
\end{cases}  
\end{equation} 

The verifier, who knows the Clifford circuit $C$, can compute a tableau description of $|\psi_C\rangle$ and hence efficiently test whether a given bitstring $z$ belongs to $S_C$. Equivalently, the support can be described as the solution space of a system of linear equations over $\mathbb{F}_2$, 
\begin{equation}\label{eq:app_affine_support} 
S_C = \{z\in \mathbb{F}_2^n : A z = b\}, 
\end{equation} 
where $A$ is a binary matrix and $b$ is a binary vector. If $\mathrm{rank}(A)=n-r$, then the solution space has dimension $r$ and size $2^r$. Membership in $S_C$ can therefore be checked by evaluating the linear constraints in Eq.~\eqref{eq:app_affine_support}. 

\subsection{Definition of the SSB statistic} 

For a bitstring $z\in\{0,1\}^n$, define the support indicator 
\begin{equation}\label{eq:app_support_indicator} 
\chi_C(z) = 
\begin{cases} 
1, & z\in S_C,\\ 
0, & z\notin S_C. 
\end{cases}  
\end{equation} 
Given $N$ measured samples $\{z_i\}_{i=1}^N$, the SSB statistic is 
\begin{equation}\label{eq:app_ssb_statistic}  
s = \frac{1}{N}\sum_{i=1}^{N}\chi_C(z_i).
\end{equation} 
Thus $s$ is the empirical probability mass that the sampled distribution assigns to the stabilizer support $S_C$. For ideal sampling from the Clifford output distribution $p_C$, every observed bitstring lies in $S_C$. Hence \begin{equation} 
\chi_C(z_i)=1 \quad\text{for all } i, 
\end{equation} 
and therefore 
\begin{equation}\label{eq:app_x_ideal}
s_{\rm ideal}=1.  
\end{equation} 
This gives the completeness of the test in the noiseless and exact obfuscation case. 

\subsection{Uninformed random sampling} 

Now consider an adversary that has no information about the hidden Clifford support $S_C$ and instead outputs uniformly random bitstrings from $\{0,1\}^n$. Let $Z$ be a uniformly random bitstring. Then 
\begin{equation}\label{eq:app_random_support_probability} 
\Pr[Z\in S_C] = \frac{|S_C|}{2^n} = \frac{2^r}{2^n} = 2^{r-n}.  
\end{equation} 
Therefore the random variable $\chi_C(Z)$ is Bernoulli distributed with success probability 
\begin{equation} 
q_r = 2^{r-n}. 
\end{equation} 

For $N$ independent uniformly random samples, 
\begin{equation} 
Ns = \sum_{i=1}^N \chi_C(Z_i) 
\end{equation} 
is binomially distributed, 
\begin{equation}\label{eq:app_binomial_random} 
Ns \sim \mathrm{Binomial}(N,q_r).  
\end{equation} 

It follows immediately that 
\begin{equation} 
\mathbb{E}_{\rm rand}[s] = q_r = 2^{r-n}, \label{eq:app_random_mean} 
\end{equation} 
and 
\begin{equation}\label{eq:app_random_variance} 
\mathrm{Var}_{\rm rand}(s) = \frac{q_r(1-q_r)}{N} = \frac{2^{r-n}\left(1-2^{r-n}\right)}{N}.  
\end{equation} 

When $r\ll n$, the random baseline is exponentially small. Even when $r$ is close to $n$, the variance decreases as $1/N$, so increasing the number of samples makes the empirical value of $s$ sharply concentrated around its mean. For example, if $r=n-1$, then $\mathbb{E}_{\rm rand}[s]=1/2$, while $s_{\rm ideal}=1$, and the standard deviation is $1/(2\sqrt{N})$. 

\subsection{Depolarizing noise} 

We next evaluate the SSB under a simple global depolarizing model. Suppose that the experimentally sampled distribution is a mixture of the ideal distribution and a uniform distribution,
\begin{equation}\label{eq:app_depolarizing_distribution}  
p_\eta(z) = (1-\eta)p_C(z) + \eta\,2^{-n}, \qquad 0\leq \eta \leq 1, 
\end{equation} 
where $\eta$ is the effective depolarizing strength. Equivalently, with probability $1-\eta$ the device samples from the ideal Clifford distribution, and with probability $\eta$ the output is replaced by a uniformly random bitstring. The expected SSB score is the total probability mass of $p_\eta$ on $S_C$: 
\begin{align} 
\mathbb{E}_\eta[s] 
&= \sum_{z\in\{0,1\}^n} p_\eta(z)\chi_C(z) \nonumber\\ 
&= \sum_{z\in S_C} \left[ (1-\eta)p_C(z) + \eta\,2^{-n} \right]. \end{align} 
Using $\sum_{z\in S_C}p_C(z)=1$ and $|S_C|=2^r$, we obtain 
\begin{equation}\label{eq:app_depolarizing_mean} 
\mathbb{E}_\eta[s] = (1-\eta) + \eta\,2^{r-n}.  
\end{equation} 

Thus depolarizing noise linearly interpolates between the ideal value $s=1$ and the uninformed-random baseline $2^{r-n}$. For the variance, each sample contributes a Bernoulli random variable $\chi_C(Z)$ with success probability 
\begin{equation} 
q_\eta = (1-\eta) + \eta\,2^{r-n}. 
\end{equation} 
Therefore 
\begin{equation} 
Ns \sim \mathrm{Binomial}(N,q_\eta), 
\end{equation} 
and hence 
\begin{equation}\label{eq:app_depolarizing_variance}  
\mathrm{Var}_\eta(s) = \frac{q_\eta(1-q_\eta)}{N}, \qquad q_\eta = (1-\eta)+\eta\,2^{r-n}. 
\end{equation} 

More generally, for any experimental output distribution $p_{\rm exp}$, the SSB statistic estimates the total probability mass assigned to the Clifford support, 
\begin{equation} 
P_{\rm exp}(S_C) = \sum_{z\in S_C}p_{\rm exp}(z). 
\end{equation} 
In that case, 
\begin{equation}\label{eq:app_general_ssb_variance} 
\mathbb{E}[s]=P_{\rm exp}(S_C), \qquad \mathrm{Var}(s) = \frac{P_{\rm exp}(S_C)\left[1-P_{\rm exp}(S_C)\right]}{N}.  
\end{equation} 

\section{Unitary Stabilizer Nullity}\label{app:unitary_stabilizer_nullity} 

In this appendix we briefly introduce the unitary stabilizer nullity used in the main text to lower bound the $T$-gate count of the obfuscated circuits. The discussion follows Ref.~\cite{Jiang2023lower}. 

Let $\overline{\mathcal{P}}_n$ denote the $n$-qubit Pauli group modulo global phases. This quotient Pauli group can be identified with a $2n$-dimensional vector space over $\mathbb{F}_2$, where each Pauli operator is represented by its binary $X$ and $Z$ components. Clifford unitaries are precisely those unitaries that normalize the Pauli group: for every $P\in \overline{\mathcal{P}}_n$, 
\begin{equation} 
C P C^\dagger \in \overline{\mathcal{P}}_n . 
\end{equation} 

A non-Clifford unitary, by contrast, maps at least some Pauli operators to non-Pauli operators under conjugation. For an arbitrary $n$-qubit unitary $U$, define its unitary stabilizer group as the subgroup of Pauli operators that remain Pauli under conjugation by $U$: 
\begin{equation}\label{eq:app_unitary_stabilizer_group} 
\mathcal{P}_U = \left\{ P\in \overline{\mathcal{P}}_n : U P U^\dagger \in \overline{\mathcal{P}}_n \right\}.  
\end{equation} 
Equivalently, $\mathcal{P}_U$ is the largest subgroup of the quotient Pauli group on which the adjoint action of $U$ behaves Clifford-like. Since $\overline{\mathcal{P}}_n$ is a $2n$-dimensional binary vector space, we may define the dimension of $\mathcal{P}_U$ as the maximum number of independent Pauli generators contained in this subgroup. The unitary stabilizer nullity of $U$ is then defined as the codimension of $\mathcal{P}_U$ inside the full quotient Pauli group: 
\begin{equation}\label{eq:app_unitary_stabilizer_nullity} 
\nu(U) = 2n - \dim(\mathcal{P}_U).  
\end{equation} 
This quantity vanishes for Clifford unitaries, because a Clifford unitary maps every Pauli operator to another Pauli operator and hence $\dim(\mathcal{P}_U)=2n$. Conversely, a nonzero value of $\nu(U)$ certifies that $U$ has non-Clifford action on the Pauli group. 

The relevance of $\nu(U)$ to fault-tolerant gate synthesis is that it lower bounds the number of $T$ gates needed to implement $U$, up to Clifford gates. More precisely, if $T_{\rm opt}(U)$ denotes the minimum number of $T$ gates over all Clifford+$T$ circuits implementing $U$, then Ref.~\cite{Jiang2023lower} proves 
\begin{equation}\label{eq:app_t_count_lower_bound} 
T_{\rm opt}(U) \geq \nu(U).  
\end{equation} 
Thus the unitary stabilizer nullity provides an efficiently computable certificate of non-Clifford resource: a large value of $\nu(U)$ rules out any Clifford+$T$ implementation with fewer than $\nu(U)$ $T$ gates. The bound should be interpreted as a lower bound rather than an exact synthesis cost. If $T_{\rm circ}(U)$ is the number of $T$ gates appearing in a particular circuit representation of $U$, and $T_{\rm opt}(U)$ is the optimal $T$ count over all equivalent representations, then \begin{equation}\label{eq:app_t_count_hierarchy} 
\nu(U) \leq T_{\rm opt}(U) \leq T_{\rm circ}(U).  
\end{equation} 

In the main text, we use $T_{\rm lower} \equiv \nu(U_{\rm obf})$ to obtain a rigorous lower bound on the non-Clifford resource generated by imperfect patch updating. Even when circuit simplification tools reduce the explicit $T$ count of a given compiled representation, Eq.~\eqref{eq:app_t_count_lower_bound} guarantees that no equivalent Clifford+$T$ circuit can have fewer than $\nu(U_{\rm obf})$ $T$ gates.

\end{document}